\documentclass[copyright,creativecommons]{eptcs}
\providecommand{\event}{ML 2015} % Name of the event you are submitting to
\usepackage{breakurl}             % Not needed if you use pdflatex only.
\title{GADTs and Exhaustiveness: Looking for the Impossible}
\author{Jacques Garrigue
\institute{Nagoya University Graduate School of Mathematics}
\email{garrigue@math.nagoya-u.ac.jp}
\and Jacques Le Normand
\email{rathereasy@gmail.com}}
\def\titlerunning{GADTs and Exhaustiveness}
\def\authorrunning{J. Garrigue \& J. Le Normand}

\usepackage{alltt}
\def\eg.{\textit{e.g.}}
\def\ie.{\textit{i.e.}}
\let\kw\texttt
\newcommand\kd[1]{{\bf#1}}

\newenvironment{program}
{\begin{list}{}{\small\leftmargin=2ex\topsep=1ex}%
 \item{}\begin{alltt}}
{\end{alltt}\end{list}}

\let\tv=\alpha
\let\tw=\beta
\let\t=\tau
\let\ra=\rightarrow
\let\th=\vdash
\let\type=\textit
\newcommand\cmt[1]{\hfill{(* #1 *)}}

\begin{document}
\maketitle

\begin{abstract}
Sound exhaustiveness checking of pattern-matching is an essential
feature of functional programming languages, and OCaml supports it for
GADTs.
However this check is incomplete, in that it may fail to detect that a
pattern can match no concrete value.
In this paper we show that this problem is actually undecidable,
but that we can strengthen the exhaustiveness and redundancy checks so
that they cover more practical cases.
The new algorithm relies on a clever modification of type inference for
patterns.
\end{abstract}

\section{Introduction}
Exhaustiveness and redundancy checks of pattern-matching are
important features of functional programming
languages~\cite{Maranget2007}. They let
programmers write and maintain software more efficiently and with
fewer bugs. In OCaml, they even allow extra optimizations in the
pattern-matching compiler~\cite{LeFessant2001icfp}\footnote{
  The propagation of exhaustiveness information to pattern matching
  compilation in OCaml was originally introduced by Jacques Garrigue,
  to allow better compilation of polymorphic variants~\cite{Garrigue98}.}.

In OCaml 4.00~\cite{OCaml400}, pattern-matching was extended to
accommodate GADTs~\cite{Garrigue2011}.
GADT is a shorthand for Generalized Algebraic Data
Types~\cite{AugustssonSilly94,Xi2003popl,Cheney2003}
which denotes an extension of Algebraic Data Types where the
definition may contain type equalities specific to a branch.
This extension clearly has an impact on exhaustiveness and redundancy
checking.
However, while the question of exhaustiveness of pattern-matching has
been studied extensively in the context of dependent type
systems~\cite{Coquand92,McBride99phd,Xi99padl,Schurmann2003,Dunfield2007phd,Norell2007phd,Oury2007pplv},
the more restricted case of GADTs used in practical functional
programming languages was not well explored.
For instance GHC, which already had GADTs since 2005, didn't support
them in either check~\cite{Peyton2006}.
Ironically, even Xi, who had already shown how types could be used to
prune some cases in his work on Dependent ML~\cite{Xi99padl}, didn't
try to use that in the context of GADTs~\cite{Xi2003popl}.
One reason might be that the coverage check for dependently typed
pattern-matching is known to be undecidable in
general~\cite{McBride99phd}, and as such the road was potentially slippery.

For this same reason, we settled for a minimal modification of the
exhaustiveness check, which had to be clearly sound, but didn't
attempt to be complete.
The basic idea was to collect the missing cases from a
pattern-matching, and then use type inference to prune the
impossible ones.
While this limited modification already proved to be ambitious, as was
shown by a number of bugs in corner cases that had to be fixed, it
also appeared to be insufficient, as some users encountered spurious
warnings on code whose exhaustiveness was clear enough.
Moreover, modifying only the exhaustiveness check created an asymmetry
with the redundancy check.

In this paper we both describe the original approach, including what we had
to do to ensure its soundness, and a new improved algorithm, which
supports both exhaustiveness and redundancy checks, and is able to
prove exhaustiveness in more cases. The basic idea behind this new
algorithm is to see both checks as trying to prove that a type is not
inhabited, which can be done by turning type inference into a proof
search algorithm.
Unfortunately, a quick look a the search space, which exactly matches
resolution for Horn clauses, is enough to show that the exhaustiveness
and redundancy problems are undecidable even when limited to GADTs.
This means that this proof search must be restricted so that its
termination can be enforced in a predictable way.

The idea of using a standard non-determinism technique to turn type
inference into a proof search applicable to exhaustiveness and
redundancy checks is the main contribution of this paper.
We complete it with some basic heuristics, and syntactic additions
to the language that the programmer can use to provide some guidance
to the exhaustiveness checker in its proof search.

This paper is organized as follows. In section~\ref{sec:exhaustiveness},
we explain what is the exhaustiveness problem for GADTs through
examples. In section~\ref{sec:implementation}, we describe the
original implementation, and how we needed to introduce a new notion
of type compatibility to make it sound. In
section~\ref{sec:backtracking} we describe the new proof search
algorithm. In section~\ref{sec:undecidability} we show that
the problem is undecidable for GADTs, and discuss practical heuristics
and language features to accommodate this.
Finally, in section~\ref{sec:redundancy}, we show how the redundancy
check can be integrated in this picture, and describe the concrete
semantics implemented in OCaml 4.03.

\section{GADTs and exhaustiveness}
\label{sec:exhaustiveness}

Checking the exhaustiveness of pattern-matching is a difficult
problem. Technically, it is about checking whether there are values of
the matched type that are not covered by the cases of the
pattern-matching.
There are well-known techniques to handle this problem for algebraic
data types~\cite{Maranget2007}, but they do not attempt to tackle
semantical questions, such as whether such a value can be built or
not.
For instance consider the following example:
\begin{program}
\kd{type} empty = \{e : 'a. 'a\}
\kd{let} f : empty option -> unit = \kd{function} None -> ()
\end{program}
Since there is no way to build a value of type {\tt empty} (at least
in a type-safe call-by-value language), this match
is actually exhaustive, but the checker will still report a missing
{\tt Some} case.

For normal types, this limitation does not matter: why would one
intentionally introduce an empty type? However, in the case of GADTs,
the problem becomes acute.
\begin{program}
\kd{type} _ t =
  | Int : int t
  | Bool : bool t

\kd{let} f : \kd{type} a. a t -> a = \kd{function}
  | Int -> 1
  | Bool -> true

\kd{let} g1 : \kd{type} a. a t -> a = \kd{function}
  | Int -> 1
\type{Warning 8: this pattern-matching is not exhaustive.
Here is an example of a value that is not matched:
Bool}

\kd{let} g2 : int t -> int = \kd{function}
  | Int -> 1

\kd{let} h : \kd{type} a. a t -> a t -> bool = \kd{fun} x y ->
  \kd{match} x, y \kd{with}
  | Int, Int -> true
  | Bool, Bool -> true
\end{program}
Function {\tt f} is a classical GADT function, where different
branches instantiate the type parameter differently.
It is clearly exhaustive.
If we just remove the second case without changing the type
annotation, in function {\tt g1}, the pattern-matching becomes
incomplete, and a missing case {\tt Bool} should be reported.
However, in function {\tt g2}, we change the type annotation to
restrict its input to be of type {\tt int t}, which is 
incompatible with the constructor {\tt Bool}, so that the
only valid input is {\tt Int}, making it exhaustive.
Function {\tt h}, is also exhaustive, because it requires {\tt x} and {\tt
  y} to have the same type.
Examples {\tt g2} and {\tt h} have useful instances, and we want them
to be recognized as exhaustive, but this requires the exhaustiveness
check to take type information into account.

\section{First implementation}
\label{sec:implementation}
When we first implemented type inference for GADTs in
OCaml~\cite{Garrigue2011},
we did not know for certain that the exhaustiveness problem was
undecidable, but it seemed highly unlikely that there was a simple and
complete algorithm to solve it. Our strategy was to find a simple
conservative algorithm that was easily understandable and that would
catch all the potential bugs for the programmer.

\subsection{Algorithm}
Our first algorithm took the approach of using the original
exhaustiveness check to extract a list of missing patterns, and then
rely on the type checker to prove that they are not inhabited. For this, we
introduced the following two changes to the exhaustiveness checker.
\begin{enumerate}
\item The original exhaustiveness algorithm did return a
  counter example when a pattern-matching was not complete, but this
  example did not cover all the missing cases.
  For instance, in the previous example {\tt h}, the exhaustiveness
  checker would only return {\tt Int, Bool} as a missing pattern,
  while we also needed to check  that {\tt Bool, Int} is an invalid
  pattern to remove any possible doubt that {\tt h} is actually
  exhaustive.
  Consequently, we modified the exhaustiveness checker so that it
  would return a complete set of missing patterns.
% Note that the patterns in this set do not contain or-patterns.
\item Each pattern in the missing set is then fed to the type checker,
  to detect whether it is typeable or not. Untypeable patterns cannot be
  matched by any value, so it is safe to ignore them.
  If no pattern remains, the pattern-matching is deemed exhaustive,
  otherwise the or-pattern of all the remaining patterns is returned as
  an exhaustive counter-example.
  Note that when typing patterns, one needs to relax unification, as
  we will see in the next subsection.
\end{enumerate}

\iffalse
It is worth mentioning that, while we (re-)discovered this algorithm
on our own, and we provide only an informal description, Duncan for
instance proposed a type system strong enough to check exhaustiveness
in the same way~\cite{Dunfield2007phd}.
\fi

\subsection{Incompleteness}
\label{sec:incompleteness}
This approach seemed sufficient at first, as almost all exhaustive
pattern-matchings were detected as such.
However, enthusiastic GADT users were more clever than
the checker, and they got exhaustiveness warnings where none should
be~\cite{PR6437}. The problem there involved using variables from a
finite set to access a finite length context, but the essential cause
can be seen in the following example.
\begin{program}
\kd{type} \_ is\_int = IsInt : int is\_int

\kd{let} h2 : \kd{type} a. a t * a is\_int -> bool = \kd{function}
  | Int, IsInt -> true
\end{program}
The generated counter-example here is {\tt (Bool, \_)}.
As this pattern appears to be typeable, a warning was emitted.
This warning is of course spurious, since the only value which can
replace the wildcard is {\tt IsInt}, and its type would be
incompatible with {\tt Bool}. We will see later that this replacement
of wildcards by the concrete cases from the type definition can be
done systematically.
As a side remark, one may notice that reversing the components of the
pair in this example makes exhaustiveness detection easier: then the
only counter-example becomes {\tt (IsInt, Bool)}, which is clearly
impossible. However, this is no sufficient reason to just ignore the
problem.

\subsection{Type compatibility}
\label{sec:compatibility}
While reusing the type checker to detect non matchable patterns seems a
good idea, it relies on the following fact:
\newtheorem{fact}{Fact}
\begin{fact}
If a pattern $p$ matches a value $v$, and $v$ can be given some
type $\t$, and $p$ does not contain or-patterns, then $p$ can also be
given type $\t$.
\end{fact}
A similar lemma appears in~\cite{Xi99padl} for instance.
This seems reasonable: a pattern is built like a value,
except that some parts may be wild cards, which can be replaced by any
value of the same type, so that this fact can be proved by an easy
induction.

However, one must be careful that typeability in ML is a function of
the context. Concretely, the ML module system allows one to hide type
equalities, so that it is for instance impossible to know if two
abstract types exported from other modules are equal or not. Take, for
example:

\begin{program}
\kd{type} (_, _) cmp =
 | Eq : ('a, 'a) cmp
 | Any: ('a, 'b) cmp

\kd{module} A : \kd{sig} \kd{type} a \kd{type} b \kd{val} eq : (a, b) cmp \kd{end}
  = \kd{struct} \kd{type} a \kd{type} b = a \kd{let} eq = Eq \kd{end}

\kd{let} f : (A.a, A.b) cmp -> unit = \kd{function} Any -> ()
\type{Warning 8: this pattern-matching is not exhaustive.
Here is an example of a value that is not matched:
Eq}
\end{program}

\begin{figure}
\[
\th_E \tv \simeq \tw
\hspace{5ex}
\frac{\mathrm{type}~(\vec\tv)t \in E \quad \t'\neq(\vec\theta')t}
     {\th_E (\vec \t)t = \t'}
\hspace{5ex}
\frac{\th_E \t_i \simeq \t'_i  ~~(i\in \mathrm{Inj}_E(t))}
     {\th_E (\t_1,\dots,\t_n)t \simeq (\t'_1,\dots,\t'_n)t}
%\frac{\mathrm{type}~(\vec\tv)t = \t_t \in E \quad \th_E (\vec\t)t \simeq \t'}
%     {\th_E [\vec\t/\vec\tv]\t_t = \t'}
\]
\[
\th_E \t \simeq \t
\hspace{5ex}
\frac{\th_E \t_1 \simeq \t_2}{\th_E \t_2 \simeq \t_1}
\hspace{5ex}
\frac{\th_E \t_1 \simeq \t'_1 \quad \th_E \t_2 \simeq \t'_2}
     {\th_E \t_1 \ra \t_2 \simeq \t'_1 \ra \t'_2}
\hspace{5ex}
\frac{\th_E \t_1 \simeq \t'_1 \quad \th_E \t_2 \simeq \t'_2}
     {\th_E \t_1 \times \t_2 \simeq \t'_1 \times \t'_2}
\]
\[
\begin{array}c
\mathrm{type}~(\vec\tv)~t = C_1\mbox{ of }\theta_1 \mid \dots \mid
  C_n\mbox{ of }\theta_n
\quad
\mathrm{type}~(\vec\tv')~t' = C_1\mbox{ of }\theta'_1 \mid \dots \mid
  C_n\mbox{ of }\theta'_n \\
t \neq t' \quad
\th_E \theta_i \simeq \theta'_i ~~(1\leq i \leq n) \quad
\th_E \t_i \simeq \t'_i \\ \hline
{\th_E (\t_1,\dots,\t_n)t \simeq (\t'_1,\dots,\t'_n)t'}
\end{array}
\]
\caption{Compatibility relation}
\label{fig:compat}
\end{figure}

Since there is no relation between \kw{a} and \kw{b} in the signature
of \kw{A}, {\tt Eq} cannot be given the type {\tt (A.a, A.b) cmp} outside of
module {\tt A}, which might lead one to think that the function {\tt
  f} is exhaustive.
However, inside \kw{A} we can use the type equality {\tt a = b} to
create such a value and export it.

The exhaustiveness checker should detect that {\tt f} is not
exhaustive.
Of course, it cannot do it by searching the whole typing environment
for a way to build such a value, as this would clearly be undecidable,
and we need a conservative answer.
Rather, we change the definition of typeability for patterns, so that
{\tt Eq}, as a pattern, can be given the type {\tt (A.a, A.b) cmp}.

In general, this means that we allow typing patterns
assuming extra type equalities for abstract types.
Technically, a new compatibility relation on types is introduced, and when
the type checker unifies indices of GADTs (the type
parameters of a GADT constructor) during the typing of patterns, it
refers to  this relation for non-unifiable type
constructors appearing inside these indices, rather than immediately
raising a unification error.
The compatibility relation (for only some features of the language) is
the smallest relation including the rules of
Figure~\ref{fig:compat}.
For simplicity, this relation doesn't try to be too restrictive, and
for instance it doesn't distinguish between different type variables.
More importantly, the second rule expresses our assumption
that abstract types are compatible with all other types.
As a result of these 2 rules, this relation is reflexive and symmetric,
but not transitive.
Additionally, the next rule, concerning identical
type constructors, says that we shall only check compatibility for
injective parameters, \ie.~parameters whose equality can be deduced
from the equality of the resulting types.
Finally the last rule says that two data types are compatible if their
data constructors have the same names (and order), and the types of
their arguments are compatible.

In our example above, when typing the missing case {\tt Eq}, the
compatibility relation says that {\tt A.a} and {\tt A.b} are
compatible.
This compatibility relation makes the type checker far more permissive
with GADT indices inside patterns than inside expressions. 
This is fine because, in doubt, it is safe to allow possibly
impossible patterns and to reject potentially unsafe expressions.

Note that since we use exactly the same function to type-check
patterns and to check exhaustiveness, we have the nice property that
if the exhaustiveness check reports a missing pattern, then type
checking will always allow it.
This is to contrast with what used to happen in GHC before version 8.0,
as the exhaustiveness checker could tell you that some case was missing,
but adding it would cause a type error, leaving the programmer with
the only option of adding a catch all case.
Thanks to recent work~\cite{Karachalias2015}, GHC 8.0 has the
same property, by using the same typechecking-based pruning; note that
since Haskell does not allow the same kind of type abstraction as ML,
they do not need our compatibility relation.

\iffalse
The GHC compiler has one very confusing behaviour: the exhaustiveness
checker may flag a pattern as missing, yet the type checker would not
allow you to add it. It's a kind of catch 22.
We do not have this problem because  we use exactly the same function
to type-check patterns and to check exhaustiveness: if the
exhaustiveness check reports a missing pattern, then type checking
will always allow it.
Thanks to recent work~\cite{Karachalias2015}, GHC 8.0 should have the
same property, by using the same pruning approach; note that since
Haskell does not allow the same kind of type abstraction as ML, they
do not need our compatibility relation.
\fi

\section{Splitting and backtracking}
\label{sec:backtracking}
While our original approach seemed mostly satisfactory, there are
cases where it fails.
We have already seen such an example in Section~\ref{sec:incompleteness}.
Here is an even simpler one.
\begin{program}
\kd{let} deep : char t option -> char = \kd{function} None -> 'c'
\end{program}
Since {\tt t} is only defined for {\tt int} and {\tt bool}, {\tt char
  t} is actually the empty type, \ie.~there are no values of the form
{\tt Some \_} at type {\tt char t option}. However, there is no order
to change, and to see that {\tt char t} is empty one needs to split
{\tt \_} into its different cases, and check them
separately.
This gives us the following two patterns:
\begin{program}
Some Int
Some Bool
\end{program}
Then we can call the type checker as before, to verify that they are
incompatible with the given type.

We just happened to rediscover in a very natural way the idea of {\em
  context splitting}, originally introduced by Coquand for doing
pattern matching on dependent types~\cite{Coquand92}.

Note that as soon as we start to do deeper case analysis, the approach
switches from just checking whether a pattern has some type to checking
whether a particular type is inhabited by terms of a certain form.
Here are a few more examples of the same kind, by order of
difficulty.
\begin{program}
\kd{type} zero = Zero
\kd{type} _ succ = Succ

\kd{type} (_,_,_) plus =
  | Plus0 : (zero, 'a, 'a) plus
  | PlusS : ('a, 'b, 'c) plus -> ('a succ, 'b, 'c succ) plus

\kd{let} trivial : (zero succ, zero, zero) plus option -> bool
  = \kd{function} None -> false

\kd{let} easy : (zero, zero succ, zero) plus option -> bool
  = \kd{function} None -> false

\kd{let} harder : (zero succ, zero succ, zero succ) plus option -> bool
  = \kd{function} None -> false
\end{program}
{\tt zero} and {\tt succ} encode type level natural numbers.
{\tt plus} is the Peano version of addition, in relational form;
namely there is a term {\tt ($a,b,c$) plus} if and only if $a+b=c$.
{\tt trivial} can be easily checked, as {\tt (zero succ, zero, zero)}
does not match either of {\tt Plus0} and {\tt PlusS}.
{\tt easy} is a bit more difficult, as it seems to match {\tt Plus0},
but unification between {\tt zero succ} and {\tt zero} fails later.
For {\tt harder}, unification with {\tt PlusS} succeeds, however the
argument becomes {\tt (zero, zero succ, zero) plus}, which was
inferred empty in {\tt easy}.

In {\tt deep}, {\tt trivial} and {\tt easy} it is sufficient to
split the first {\tt \_} according to its inferred type.
However, {\tt harder} requires to infer the type of the argument of
the GADT constructor {\tt PlusS} in order to split it once more.

Another interesting case is when there is a dependency between
components of a tuple.
\begin{program}
\kd{let} inv_zero : \kd{type} a b c d. (a,b,c) plus -> (c,d,zero) plus -> bool
  = \kd{fun} p1 p2 ->
    \kd{match} p1, p2 \kd{with}
    | Plus0, Plus0 -> true
\end{program}
Here the extra patterns coming from the basic exhaustiveness algorithm
are:
\begin{program}
Plus0, PlusS _
PlusS _, _
\end{program}
While the first pattern is clearly empty, the second one is typeable if
one does not split the second {\tt \_}. However, to do that we would
need to first infer the type of the second component of the pair,
which depends on the freshly generated first component. In this case
again, typing patterns (for checking emptiness) and splitting
wildcards must be interleaved. From a more theoretical point of view,
this is the reason why Coquand used the name {\em context splitting}, rather
than just {\em pattern splitting}. Previous choices change the
splitting we can do next.

The solution to this inter-dependence is to actually do all of these
simultaneously.
Namely, we modified the recursive {\tt type\_pat},
which is the main function for typechecking patterns, in order to turn
it into a proof-searching function.
The basic idea is to make it non-deterministic.
However, since this function uses side-effecting unification,
returning a list of results would not be easy. Rather we converted it to 
continuation passing style, using backtracking to cancel
unification where needed. In particular, it is sufficient to split
wild cards into or-patterns\footnote{
{\em or-patterns} are patterns of the form $\it pat_1 \mid pat_2$,
which catch the union of $\it pat_1$ and $\it pat_2$. Since we are
splitting wild cards, in this case they do not contain variables.},
as they are then interpreted in a
non-deterministic way, allowing to check all combinations.
\begin{program}
(* mode is Check or Type, k is the continuation *)
\kd{let rec} type_pat mode env spat expected_ty k =  
  \kd{match} spat.ppat_desc \kd{with}
  | Ppat_any -> (* wild card *)
    \kd{if} mode = Check && is_gadt expected_ty \kd{then}
      type_pat mode env
        (explode_pat !env expected_ty)
        expected_ty k
    \kd{else} k (mkpat Tpat_any expected_ty)
  | Ppat_or (sp1, sp2) -> (* or pattern *)
    \kd{if} mode = Check \kd{then}
      \kd{let} state = save_state env \kd{in}
      \kd{try} type_pat mode env sp1 expected_ty k
      \kd{with} exn ->
        set_state state env;
        type_pat mode env sp2 expected_ty k
    \kd{else}
    (* old code *)
  | Ppat_pair (sp1, sp2) -> (* pair pattern *)
    \kd{let} ty1, ty2 = filter_pair env expected_ty \kd{in}
    type_pat mode env sp1 ty1 (\kd{fun} p1 ->
    type_pat mode env sp2 ty2 (\kd{fun} p2 ->
      k (mkpat (Tpat_pair (p1,p2)) expected_ty)))
  |  ... (* other cases in CPS *)
\end{program}
As you can see, this modified {\tt type\_pat} function\footnote{The code
  is part of OCaml source code since version 4.03, and can be found in
  the online repository:
  \url{https://github.com/ocaml/ocaml/blob/trunk/typing/typecore.ml}.}
has a special
handling for {\tt Ppat\_or} in {\tt Check} mode, which behaves as though
the whole pattern were duplicated, with the or-pattern replaced by
either of its sub-patterns.
This means that, rather than calling {\tt type\_pat} individually on
each missing pattern, it becomes possible to call it directly on an
or-pattern combining all the missing cases, and it is fine to leave
some or-patterns deep inside. This makes generating the patterns for
missing cases simpler, but more importantly this allows to introduce
such or-patterns on the fly by splitting wild cards while searching
for a typeable case, as you can see in the handling of {\tt Ppat\_any}.
For the remaining cases, such as {\tt Ppat\_pair}, one just has to use
continuation passing style, to allow the remainder of the code to be
called several times, instantiated with different branches of
or-patterns.

\iffalse
While we believe that our approach of introducing non-determinism in
an already existing type checking algorithm in order to check
(potential) inhabitation is new, the idea of having the type system
itself ensure the exhaustiveness of pattern-matching is not new.
For instance, Dunfield~\cite{Dunfield2007phd} defined such a type system
by introducing a notion of pattern subtraction, and rules allowing to
check that the remaining patterns cannot match any value.
\fi

One should not confuse this use of type checking inside the
exhaustiveness check, with other more theoretical works where the
check is integrated into type
checking~\cite{Dunfield2007phd,Krishnaswami2009popl}.
There might be some interesting connections, but here we are still
relying on an independent exhaustiveness analysis to produce the
missing patterns.

\section{Undecidability and heuristics}
\label{sec:undecidability}

\begin{figure}[t]
\begin{program}
\kd{type} s0 \kd{and} fin \cmt{states}
\kd{type} c0 \kd{and} blank \cmt{symbols}
\kd{type} left \kd{and} right \cmt{directions}
\kd{type} endt \cmt{end of tape}

\kd{type} ('st_in, 'head_in, 'st_out, 'head_out, 'lr) transition =
  | Tr1 : (s0, blank, s1, c0, left) transition
  | Tr2 : (s1, blank, s0, c0, right) transition
  | Tr3 : ('s, c0, fin, c0, left) transition

\kd{type} ('state, 'head, 'left, 'right) eval =
  | Tm_fin       : (fin, _, _, _) eval
  | Tm_mv_left   : ('st_in, 'head_in, 'st_out, 'head_out, left) transition *
                   ('st_out, 'head_out, 'left, 'head_out * 'right) eval ->
                   ('st_in, 'head_in, 'head * 'left, 'right) eval
  | Tm_mv_right  : ('st_in, 'head_in, 'st_out, 'head_out, right) transition *
                   ('st_out, 'head_out, 'head_out * 'left, 'right) eval ->
                   ('st_in, 'head_in, 'left, 'head * 'right) eval
  | Tm_ext_left  : ('state, 'head, blank * endt, 'right) eval ->
                   ('state, 'head, endt, 'right) eval
  | Tm_ext_right : ('state, 'head, 'left, blank * endt) eval ->
                   ('state, 'head, 'left, endt) eval

\kd{type} goal = (s0, blank, endt, endt) eval

\kd{let} f : goal option -> unit = \kd{function} None -> ()
\end{program}
\caption{Encoding of the halting problem}
\label{fig:halting}
\end{figure}

The above definition of {\tt type\_pat}, in all its expressive
power, gives also a strong hint at why exhaustiveness checking of
GADTs is undecidable. A simple way to see it is that GADTs can encode
Horn clauses in a very direct way, each type definition being a
predicate, and each constructor a clause, with its arguments the
premises. For instance the type {\tt plus} defined in
Section~\ref{sec:backtracking} is actually an encoding of the following
Horn clauses, in Prolog syntax:
\begin{program}
plus(zero, X, X).
plus(succ(X), Y, succ(Z)) :- plus(X, Y, Z).
\end{program}
Reciprocally, any GADT can be encoded as a set of Horn clauses.
\begin{program}
\kd{type} _ expr =
  | Int : int -> int expr
  | Add : (int -> int -> int) expr
  | App : ('a -> 'b) expr * 'a expr -> 'b expr
\end{program}
would become:
\begin{program}
int.  \cmt{int is inhabited}
expr(int) :- int.  \cmt{Int case}
expr(int -> int -> int). \cmt{Add case}
expr(B) :- expr(A -> B), expr(A). \cmt{App case}
\end{program}
Then the {\tt type\_pat} function implements
Prolog's SLD resolution, for which counter-example generation
(\ie.~construction of a witness term) is known to be at best
semi-decidable (since resolution for Horn clauses is only
semi-decidable).

Using this view of GADTs as Horn clauses, we can also encode an
implementation of Turing machines in Prolog into GADT definitions,
so that exhaustiveness checking is equivalent to the halting problem.
This encoding is shown in Figure~\ref{fig:halting}, where {\tt
  transition} encodes the definition of the Turing machine (its valid
transitions), and {\tt eval} encodes the execution of this Turing
machine, succeeding if the final state {\tt fin} is reached.
The function {\tt f} is exhaustive if and only if the Turing machine
terminates starting from state {\tt s0} on an empty tape.

This proof of undecidability is not new: McBride already demonstrated an
inductive type encoding traces of Turing machines in
\cite[Section~6.4]{McBride99phd}, and used it as proof of the
undecidability of exhaustiveness. Similarly, Oury gave an even shorter
example encoding the Post correspondence problem~\cite{Oury2007pplv}.
In both cases, it is trivial to convert definitions so that they fit
into the scope of GADTs. What is newer is the use of Horn clauses to
make the connection between a semi-decidable implementation in Prolog
and the corresponding GADT definitions.

It is interesting to note that GHC does not have the same undecidability
problem, at least if we restrict it to lazy pattern matching.
Namely, in GHC every type is inhabited by the value {\tt undefined}.
\begin{program}
undefined :: forall a. a
undefined = undefined
\end{program}
As a result, one only has to check the remaining cases when a
constructor is actually matched on, as this forces evaluation, but not
for wild cards. This explains why Karachalias {\it et
  al.}~\cite{Karachalias2015} did not discuss this problem.
Note however that, since GHC has strictness annotations, one can
actually create situations where inhabitation by constructor terms
becomes relevant, and the exhaustiveness check becomes
undecidable.

For OCaml, this undecidability means that we have to find a good
heuristics as to where to abandon the search. Note that the complexity
is exponential in the number of wildcard patterns split.
A simple heuristics, that seems sufficient in most cases, is to only
split wildcard patterns when they do not generate multiple branches
(\ie.~tuple types or data types with a single case), or when they have
only GADT constructors (but then do not split any of the generated
subpatterns).
This means that the {\tt harder} example above would be flagged
non-exhaustive while all the other examples would be correctly identified as
exhaustive. Here is another example which would be incorrectly
flagged:
\begin{program}
\kd{type} ('a,'b) sum = Inl of 'a | Inr of 'b;;
\kd{let} deeper : (char t, char t) sum option -> char =
  \kd{function} None -> 'c'
\textit{Warning 8: this pattern-matching is not exhaustive.
Here is an example of a value that is not matched:
Some _}
\end{program}
Here the wild card points to a sum type with multiple cases, so that
the case-analysis stops there.
Even in this very limited approach, one can still exhibit an
exponential behavior:
\begin{program}
\kd{type} _ t =
  A : int t | B : bool t | C : char t | D : float t
\kd{type} (_,_,_,_) u = U : (int, int, int, int) u
\kd{let} f : \kd{type} a b c d e f g h.
   a t * b t * c t * d t * e t * f t * g t * h t
       * (a,b,c,d) u * (e,f,g,h) u -> unit =
  \kd{function} A, A, A, A, A, A, A, A, U, U -> ()
\end{program}
The above check takes about 10 seconds to exhaust all 65536 cases.
As in Prolog, one can dramatically improve performance by changing the
pattern order.

Independently of the heuristics chosen, there will always be cases
where one would like the algorithm to try harder.
An interesting approach is the concept of the {\em absurd pattern},
introduced by Agda~\cite{Norell2008}.
This pattern is a hint to the checker that no data constructor can come at
this position, which can then be proved trivially.
In Agda, a match case containing an absurd pattern has no right-hand
side.
\begin{program}
data Fin : Nat -> Set where
  fzero : \{n : Nat\} -> Fin (suc n)
  fsuc : \{n : Nat\} -> Fin n -> Fin (suc n)

magic : \{A : Set\} -> Fin zero -> A
magic ()
\end{program}
In this example, the type {\tt Fin n} denotes natural numbers smaller
than {\tt n}, so that there is no value of type {\tt Fin zero}. As a
result, one can define a function from {\tt Fin zero} to any type, by
matching on the absurd pattern {\tt ()}.

While absurd patterns have nice properties, they have also one
drawback: in Agda one has to manually split the cases, so that every
{\tt ()} occurs at a trivially impossible position. This can be pretty
verbose.

The solution we implemented in OCaml is a variant of absurd patterns,
but where we still rely on proof search to prove emptiness.
We call these {\em refutation cases}, as the syntax is denoted by a
single dot ``{\tt .}'' on the right hand side rather than by a specific
pattern.
For instance, the {\tt deeper} example above can be written
\begin{program}
\kd{let} deeper' : (char t, char t) sum option -> char = \kd{function}
  | None -> 'c'
  | Some Inl _ -> .
  | Some Inr _ -> .
\end{program}
meaning that in the last two cases the right hand side is unreachable.
Since it becomes syntactically possible to have a function with no
concrete case, we can even write the {\tt magic} function
\begin{program}
\kd{let} magic : char t -> 'a = \kd{function} _ -> . 
\end{program}
The main difference with the absurd pattern approach is that the
absurd positions are automatically inferred. Namely, we first compute
the reachable parts of the pattern, by removing the parts matched by
previous cases, and then apply the above restricted proof search to
check that the remaining pattern is empty. Since one can write more
precise patterns, it becomes possible to point down impossible
positions explicitly; in practice it suffices to write a wildcard, and
the proof search will look for the actual contradiction inside this
wildcard.
This means that the following (exhaustive) variant of {\tt harder} is
very compact:
\begin{program}
\kd{let} harder' : (zero succ, zero succ, zero succ) plus -> 'a = \kd{function} PlusS _ -> .
\end{program}

\section{Unused cases, refutation cases, and exhaustiveness}
\label{sec:redundancy}
The dual of exhaustiveness checks is the detection of unused cases.
Take, for example:
\begin{program}
\kd{let} deep' : char t option -> char = \kd{function}
  | None -> 'c'
  | Some _ -> 'd'
\end{program}
Since we added a pattern at the end of an already exhaustive match,
it is clearly redundant.

The approach is similar: after refining the pattern to keep only
subcases that are not covered by previous cases, one must check
whether they are inhabited or not.
While detecting unused cases is technically less important ---there is
no direct impact on soundness for instance--- having accurate
warnings helps the programmer reason about his program.

One can easily see that unused cases and refutation cases use exactly
the same algorithm, {\it i.e.}~if a case can be proved unused, it can
be turned into a refutation case. In OCaml 4.03, the above definition
causes the following warning:
\begin{program}
\type{Warning 56: this match case is unreachable.
Consider replacing it with a refutation case '<pat> -> .'}
\end{program}
Note that we only suggest a refutation case if types are needed to
prove emptiness.
Of course, in the same way that we cannot guarantee that all
counter-examples of exhaustiveness are really inhabited,
we cannot hope to detect all unused cases, but at least we have here
some form of symmetry.

Another consequence of this symmetry is that refutation cases are
actually handled by the redundancy check rather than the
exhaustiveness check.
Namely, the algorithm implemented in OCaml 4.03 works as follows:
\begin{enumerate}
\item For each case, compute its {\em residual}, \ie.~the intersection of
  the pattern with the complement of the previous cases.
\item Check whether this residual is inhabited or not, using {\tt
  type\_pat}, which implements the above heuristics of splitting wild
  cards only once.
\item If this is a refutation case, and the residual could not be
  proved empty,  emit an error as the refutation failed.
\item If this is a concrete case, and the residual is empty
  emit a warning as this case is unused.
\item After processing all the cases, compute the or-pattern of the
  missing cases, which is actually the same as computing the residual
  of an extra catch-all case.
\item If the pattern-matching contained only one case, use {\tt
  type\_pat} with the same heuristics to check that this final residual
  is empty. If there are more than one cases, use {\tt type\_pat}
  without splitting. In both cases, emit a non-exhaustiveness warning
  if the residual cannot be proved empty.
\end{enumerate}
The choice of using a different heuristics for exhaustiveness when there
are more than one case may seem surprising, but it appears that there
is a significant difference in cost for usual code.
\begin{table}[h]
%\vspace{-2ex}
\[
\begin{array}{|c|c|c|}\hline
\mbox{Exhaustiveness(1/many)} & \mbox{Redundancy} & \mbox{Time(sec)}
\\ \hline
1/1 & 1 &          7.50 \\
1/0 & 1 &        7.00 \\
0/0 & 1 &          7.00 \\
0/0 & 0 &          6.75 \\ \hline
\end{array}
\]
%\vspace{-3ex}
\caption{Compilation times for the standard library}
\label{tbl:times}
\end{table}
Table~\ref{tbl:times} shows compilation times for OCaml's standard
library, which contains both GADT-free code (most modules) and
GADT-heavy code (the Format module and its dependencies).
Of course, one can reduce cost more by never splitting wild cards, but
as discussed above this may make code more verbose. On the other hand,
disabling splitting for exhaustiveness alone is less of a problem, as
one can always add a plain refutation case ``{\tt \_ -> .}''.
The rationale for having a different behavior when there is only one
case is that some syntactic forms do not allow to add an extra case,
and the extra cost appears to be insignificant. In particular, this
allows to prove exhaustiveness for all examples here except {\tt harder}
and {\tt deeper}.

\section{Conclusion}
Exhaustiveness and redundancy checks have been an important part of the
OCaml compiler since its inception, but only recently received a major
overhaul because of the introduction of GADTs. The initial changes
were the smallest changes we could make which accommodated GADTs and
consequently the resulting algorithm had some shortcomings. By
discovering the link between exhaustiveness checks and logic programming
(Horn clauses), we were able to significantly improve the algorithm
and at the same time understand the limits of what we could possibly
do. The introduction of refutation cases to overcome these limits
actually furthers the duality of exhaustiveness and redundancy
checks, to give us a more regular system of warnings.

\section*{Acknowledgements}
We are grateful for the useful comments provided by anonymous
reviewers, and for the feedback from the OCaml community on our
experiments. This work was partially supported by JSPS Grant-in-Aid
for Scientific Research 16K00095.

\bibliographystyle{eptcs} \bibliography{gadtspm}

\end{document}